\input harvmac.tex

\def\sssec#1{$\underline{\rm #1}$}

% Something to deal with sub-sub-sections

\def\unlockat{\catcode`\@=11}
\def\lockat{\catcode`\@=12}

\unlockat
% Something to deal with sub-sub-sections

\def\newsec#1{\global\advance\secno by1\message{(\the\secno. #1)}
\global\subsecno=0\global\subsubsecno=0\eqnres@t\noindent
{\bf\the\secno. #1}
\writetoca{{\secsym} {#1}}\par\nobreak\medskip\nobreak}
\global\newcount\subsecno \global\subsecno=0
\def\subsec#1{\global\advance\subsecno by1\message{(\secsym\the\subsecno. #1)}
\ifnum\lastpenalty>9000\else\bigbreak\fi\global\subsubsecno=0
\noindent{\it\secsym\the\subsecno. #1}
\writetoca{\string\quad {\secsym\the\subsecno.} {#1}}
\par\nobreak\medskip\nobreak}
\global\newcount\subsubsecno \global\subsubsecno=0
\def\subsubsec#1{\global\advance\subsubsecno by1
\message{(\secsym\the\subsecno.\the\subsubsecno. #1)}
\ifnum\lastpenalty>9000\else\bigbreak\fi
\noindent\quad{\secsym\the\subsecno.\the\subsubsecno.}{#1}
\writetoca{\string\qquad{\secsym\the\subsecno.\the\subsubsecno.}{#1}}
\par\nobreak\medskip\nobreak}

\def\subsubseclab#1{\DefWarn#1\xdef #1{\noexpand\hyperref{}{subsubsection}%
{\secsym\the\subsecno.\the\subsubsecno}%
{\secsym\the\subsecno.\the\subsubsecno}}%
\writedef{#1\leftbracket#1}\wrlabeL{#1=#1}}% Macros for boxes
\lockat

%%%%%%%%%%%%%%%%%%%%%  Rublenye bukvy   %%%%%%%%%%%%%%%%%%%%%%%%
\def\IB{\relax\hbox{$\inbar\kern-.3em{\rm B}$}}
\def\IC{\relax\hbox{$\inbar\kern-.3em{\rm C}$}}
\def\ID{\relax\hbox{$\inbar\kern-.3em{\rm D}$}}
\def\IE{\relax\hbox{$\inbar\kern-.3em{\rm E}$}}
\def\IF{\relax\hbox{$\inbar\kern-.3em{\rm F}$}}
\def\IG{\relax\hbox{$\inbar\kern-.3em{\rm G}$}}
\def\IGa{\relax\hbox{${\rm I}\kern-.18em\Gamma$}}
\def\IH{\relax{\rm I\kern-.18em H}}
\def\IK{\relax{\rm I\kern-.18em K}}
\def\IL{\relax{\rm I\kern-.18em L}}
\def\IP{\relax{\rm I\kern-.18em P}}
\def\IR{\relax{\rm I\kern-.18em R}}
\def\IZ{\relax\ifmmode\mathchoice
{\hbox{\cmss Z\kern-.4em Z}}{\hbox{\cmss Z\kern-.4em Z}}
{\lower.9pt\hbox{\cmsss Z\kern-.4em Z}}
{\lower1.2pt\hbox{\cmsss Z\kern-.4em Z}}\else{\cmss Z\kern-.4em Z}\fi}
%%%%%%%%%%%%%%%%%%%% Calligraphic letters  %%%%%%%%%%%%%%%%%%%%%%%

\def\CS {{\cal S}}

%%%%%%%%%%%%%%%%%%%%%%%%%% Derivatives  %%%%%%%%%%%%%%%%%%%%%%%%

\def\p{\partial}

%%%%%%%%%%%%%%%%%%%% letters with bar %%%%%%%%%%%%%%%%%%%%%%%%%%

%%%%%%%%%%%%%%%%%%%%%%%%%%% Math symbols %%%%%%%%%%%%%%%%%%%%%%%

\def\Tr{\rm Tr}
\def\Id{{\rm Id}}

%%%%%%%%%%%%%%%%%%%%%%%%%%%%%%%%%%%%%%%%%%%%%%%%%%%%%%%%%%%%%%%%%
\font\manual=manfnt \def\dbend{\lower3.5pt\hbox{\manual\char127}}

\def\c{\cdot}
\def\half {{1\over 2}}
\def\ch{{\rm ch}}

\def\inbar{\,\vrule height1.5ex width.4pt depth0pt}
%%%%%%%%%%%%%%%%% Lie algebras %%%%%%%%%%%%%%%%%%%%%%

\def\lieg{{\underline{\bf g}}}

%%%%%%%%%%%%%%%%%%%%%%%%%%%%%%%%%%%%%%%%%%%%%%%%%%%%%%%%%%%%%%%%%

% Macros for boxes

\font\cmss=cmss10 \font\cmsss=cmss10 at 7pt

\def\boxit#1{\vbox{\hrule\hbox{\vrule\kern8pt
\vbox{\hbox{\kern8pt}\hbox{\vbox{#1}}\hbox{\kern8pt}}
\kern8pt\vrule}\hrule}}
\def\mathboxit#1{\vbox{\hrule\hbox{\vrule\kern8pt\vbox{\kern8pt
\hbox{$\displaystyle #1$}\kern8pt}\kern8pt\vrule}\hrule}}

%% ANOTHER SET OF MACROS

\def\inbar{\,\vrule height1.5ex width.4pt depth0pt}

\font\cmss=cmss10 \font\cmsss=cmss10 at 7pt

%REFERENCES
%

\lref\gncal{A.~ Gorsky, N.~ Nekrasov, ``Hamiltonian
systems of Calogero type and Two Dimensional Yang-Mills
Theory'', Nucl. Phys. {\bf B} (1994),  213-238, hep-th/9304047 }
\lref\gnell{A.~ Gorsky, N.~ Nekrasov, ``Elliptic Calogero-Moser System from
Two Dimensional Current Algebra'', hepth/9401021}
\lref\gnru{A.~ Gorsky, N.~ Nekrasov, ``Relativistic Calogero-Moser model as
gauged WZW theory'', Nucl.Phys. {\bf B} 436 (1995) 582-608, hep-th/9401017}
\lref\nhol{N.~ Nekrasov,    ``Holomorphic bundles and many-body
systems'', Commun. Math. Phys. 180 (1996),  587-603,  hepth/9503157}
\lref\dual{V.~ Fock, A.~ Gorsky, N.~ Nekrasov, V.~ Rubtsov,
``Duality in Many-Body Systems and Gauge Theories'', in preparation}

\lref\gmmm{
A.~ Gorsky,~ I.~ Krichever,~ A.~ Marshakov,~ A.~ Morozov,~ A.~ Mironov,~
``Integrablity and
Seiberg-Witten Exact Solution'',
hep-th/9505035,  Phys.Lett.B355: 466-474, 1995}
\lref\bbkt{O.~ Babelon, E.~ Billey, I.~ Krichever, M.~ Talon,
``Spin generalization of
the Calogero-Moser system and the Matrix KP equation'',
hepth/9411160}
\lref\kr{ I.~ Krichever, Funk. Anal. and Appl., {\bf 12}
(1978),  1, 76-78; {\bf 14} (1980), 282-290}
\lref\krz{I.~ Krichever, A.~ Zabrodin, ``Spin Generalization of
Ruijsenaars-Schneider model, Non-abelian Toda Chain and Representations
of Sklyanin Algebra'', hep-th/9505039}

%%%%%%%%%%%%%%%%%%%%%
\lref\fa{G.~ Faltings, ``A proof of Verlinde formula'',
J.Alg.Geom.{\bf 3}, (1994) }
\lref\beil{A.~ Beilinson, V.~ Drinfeld,
``Quantization of
Hitchin's Integrable System and Hecke Eigensheaves'',
{\rm A. Beilinson's lectures at IAS, fall 1994}}
\lref\be{D.~ Bernard, Nucl. Phys. {\bf B}303 (1988) 77,
{\bf B}309 (1988), 14}
\lref\c{F.~ Calogero, J.~ Math.~ Phys. {\bf 10} (1969), 2191; ibid. 2197\semi 
J.~ Math.~ Phys. {\bf 12} (1971) 419}
\lref\ch{I.~ Cherednik,
``Difference-elliptic operators and root systems'', hep-th/9410188}
\lref\eq{P.~ Etingof,
``Quantum integrable systems and
representations 
of Lie algebras'', J.~ Math.~ Phys. {\bf 36}(1995)2636-2651, hep-th/9311132}
\lref\ek{P.~ Etingof, A.~ Kirillov, Jr., ``On the affine
analogue of Jack's and Macdonald's polynomials'', Yale preprint, 1994}
\lref\fv{G.~ Felder, A.~ Varchenko, ``Integral formula for the
solutions of the elliptic KZB  equations'', Int. Math. Res. Notices,
1995, vol. 5, pp. 222-233}
\lref\ga{R.~ Garnier, ``Sur une classe de system\`es differentiels
Abeli\'ens
d\'eduits de la theorie des \'equations lin\'eares'', Rend. del Circ.
Matematice Di Palermo, {\bf 43}, vol. 4 (1919)}
\lref\gau{M.~ Gaudin, Jour. Physique, {\bf 37} (1976), 1087-1098}
\lref\iv{D.~ Ivanov,
``Knizhnik-Zamolodchikov-Bernard equations on Riemann surfaces'',
hep-th/9410091}
\lref\kks{D.~ Kazhdan, B.~ Kostant and S.~ Sternberg, Comm. on Pure and
Appl. Math., vol. {\bf XXXI}, (1978), 481-507}
\lref\lo{A.~ Losev, ``Coset construction and Bernard Equations'',
CERN-TH.6215/91}
\lref\op{M.~ Olshanetsky, A.~ Perelomov, Phys. Rep. {\bf 71} (1981), 313}
\lref\m{J.~ Moser, Adv.Math. {\bf 16} (1975), 197-220; }
\lref\rs{S.~N.~M. Ruijsenaars,
H. Schneider, Ann. of Physics {\bf 170} (1986),
 370-405}
\lref\r{S.~N.~M.~ Ruijsenaars, Comm. Math. Phys. {\bf 110}  (1987), 191-213 }
\lref\s{B.~ Sutherland, J. Math. Phys. {\bf 12} (1971), 246, ibid. 251\semi
Phys. Rev. {\bf A4} (1971) 2019; Phys. Rev. {\bf A5} (1972), 1372-1376;}
\lref\sch{L.~ Schlesinger, `` \"Uber eien Klasse
von Differentialsystemen
beliebiger Ordnung mit festen kritischen Punkten'',
Journal f\"ur die reine und angewandte Mathematik,
Bd. CXLI (1912), pp. 96-145}
\lref\er{B.~ Enriquez, V.~ Rubtsov, ``Hitchin systems, higher
Gaudin operators and {\sl r}-matrices'', alg-geom/9503010}

%%%%%%%%
\lref\gibbons{J.~ Gibbons, 
T.~ Hermsen, ``A generalization of the Calogero-Moser system'',
Physica {\bf 11D} (1984) 337}
\lref\wojcie{S. ~ Wojciechowski, Phys. Lett. A 102 (1984) 85}
\lref\opr{M.~ Olshanetsky, A.~ Perelomov, Inv. Math. {\bf 31} (1976), 93} . 
\lref\perelomov{A.~ Perelomov, Comm. Math. Phys. {\bf 63} (1978) 9}
\lref\perlprep{A.~ Perelomov, Preprint ITEP {\#} 27, 1976}
\lref\perlbook{A.~ Perelomov, ``Integrable Systems of 
Classical Mechanics and Lie Algebras'',
Moscow,  `Nauka', 1990 [in Russian] }
\lref\poly{ A.~P.~ Polychronakos, Phys.Lett.{\bf B} 266 (1991), 29}

%\draft

\Title{ \vbox{\baselineskip12pt\hbox{hep-th/9707111}
\hbox{ HUTP-97/A017}
\hbox{ ITEP-TH-16/97 }}}
{\vbox{
\centerline{On a Duality in Calogero-Moser-Sutherland Systems}
}}
\bigskip
\centerline{\sl To Vladimir Igorevich Arnol'd on his 60th birthday}
\bigskip
\centerline{N.~ Nekrasov\footnote{*}{Junior Fellow,
Harvard Society of Fellows}}

\vskip 1cm
\centerline{\it Lyman 
Laboratory of Physics,  Harvard University, Cambridge MA 02138, USA}
\centerline{and}
\centerline{\it Institute 
of Theoretical and Experimental Physics, 117259, Moscow, Russia}
\vskip 1cm
\centerline{nikita@string.harvard.edu\footnote{**}{A talk given at  
the seminar on 
symplectic
geometry at MIT in November 1996}}
\vskip 1.1cm
We point out  a map  between the dynamics of a 
non-relativistic system of $N$ particles in one dimension
interacting via the pair-wise potentials
$U_{I}(q) = {{{\nu}^{2}}\over{4R^{2}{\sin}^{2}({{q}\over{2R}})}}$ and the 
one of the particles with the pair potential 
$U_{II}(q) = {{{\nu}^{2}}\over{q^{2}}}$ and  the external potential $U_{ext} =
{{{\omega}^{2}q^{2}}\over{2}}$. The natural relation between the frequency 
$\omega$ and the radius
$R$ is: $\omega R = 1$.
\Date{April  1997}
\newsec{Formulation of the problem and the main statement}
Consider a system of indistinguishable
 particles on a circle $S^{1}_{R}$
of the radius $R$, interacting
with the pair-wise potential
\eqn\poti{
U_{I}(q) = {{{\nu}^{2}}\over{4R^{2} {\sin}^{2}
\bigl( {{q}\over{2R}} \bigr)}}
}
From the Hamiltonian point of view the system has the phase
space:
\eqn\phspi{
M_{I} = \Bigl[ T^{*}\bigl( S^{1}_{R} \bigr)^{N} \Bigr] /{\CS}_{N}
}
where $\CS_{N}$ is the $N$-th symmetric group. The coordinates
in the phase space will be denoted as
$(p_{i}, q_{i})$ where $q_{i}$ is the
angular coordinate on the circle $S^{1}_{R}$
and $p_{i}$ is the corresponding momentum.
The Hamiltonian $H_{I}$
which corresponds to \poti\ has the natural form:
\eqn\hami{
H_{I} = \sum_{i=1}^{N}
{{p_{i}^{2}}\over{2}} +
\sum_{i \neq j} U_{I}(q_{i} - q_{j})
}
This is a well-known Sutherland model \s.

The second model of our interest is the Calogero
model \c, which describes the particles on a real line $\IR$
with the Hamiltonian
\eqn\hamii{
H_{II} = \sum_{i} {{p_{i}^{2}}\over{2}} + 
{{{\omega}^{2}q_{i}^{2}}\over{2}} + \sum_{i <j} U_{II}(q_{i}- q_{j}), 
\quad U_{II}(q) =
{{{\nu}^{2}}\over{q^{2}}} }
The phase space is
\eqn\phspii{
M_{II} =\Bigl[ T^{*} {\IR}^{N} \Bigr] /{\CS}_{N}}

The systems $I$ and $II$ are integrable. Each Hamiltonian $H_{I,II}$
is contained in the commuting family of Hamiltonians
$H_{I,II}^{(k)}$ with $k = 0, \dots , N-1$. We claim that
these two integrable systems are
equivalent in the following sense:

\sssec{The \quad main \quad result.}  {\it There exists an
injective map
\eqn\mp{\pi : M_{II} \to M_{I}}
such that
\item{a.} $H_{II}^{(k)} = {\pi}^{*} H_{I}^{(k)}$
\item{b.} $H_{I} = H^{(1)}_{I}$, $H_{II} = H^{(0)}_{II}$
\item{c.} Locally $\pi$ is a symplectomorphism.
%The map $\pi$ is the action-angle map for the system $II$.
}
In physical terms, the image of the map $\pi$ consists of 
the configurations of particles
on a circle, which all have roughly speaking 
{\it positive} momenta. More precisely,
the action variables must be non-negative in appropriate normalization. 
Of course, the sign of  the
action variable is not uniquely defined. What we mean by positivity 
here is that under some choice
of the signs all the action variables must be non-negative. This condition 
defines a chamber in the
phase space $M_{I}$ which is the image of ${\pi}$. 

\newsec{Construction via Hamiltonian reduction}

Recall the Hamiltonian reductions giving rise to the
systems $I$ and $II$ respectively.
The main idea behind the construction is to realize the (classical) motion due to the Hamiltonians 
$H_{I,II}$ as a projection of the simple motion on a somewhat larger phase space \opr.

\sssec{Construction \quad of \quad  the \quad system \quad {\it I}}.
In the $I$ case (we essentially follow \kks\  here) one starts with the symplectic
manifold $X_{I} = T^{*}G \times {\IC}^{N}$ where $G = U(N)$ with the
canonical Liouville form
\eqn\cnfrmi{
\Omega_{I} = i {\Tr}  \delta \bigl( p \wedge \delta g g^{-1}  \bigr)
+{1\over{2i}} \delta v^{+} \wedge \delta v}
Here $p$ represents the cotangent vector to the group $G$. We think of it as of the {\it Hermitian} matrix.
The manifold $X_{I}$ is  acted on by $G$ (by conjugation
on the $T^{*}G$ factor and in a standard way on $\IC^{N}$). The action
is Hamiltonian with the moment map:
\eqn\momi{
\mu_{I} = p - g^{-1}pg - v \otimes v^{+}}
One performs the reduction at the central level of the moment map, i.e.
takes the manifold $\mu_{I}^{-1} ( -  \nu \cdot {\Id} )$ and takes
its quotient by $G$ (as it is invariant). Explicitly, one solves the equation
\eqn\momie{p - g^{-1}pg - v \otimes v^{+} = -\nu \Id}up to the $G$-action. The way to do it
is to fix a gauge
$$
g = {\exp} \Bigl( {i\over{R}} {\rm diag} (q_{1}, \ldots, q_{N}) \Bigr)
$$
and then solve for $p$ and $v$. One has:
$$
v_{i} = {\sqrt{\nu}}, \quad p_{ij} = R  p_{i} \delta_{ij} + {\nu} {{1- {\delta}_{ij}}\over{e^{{i(q_{i}-q_{j})}\over{R}} -1}}
$$

As a result one gets the reduced phase space:
\eqn\redi{
M_{I} = \Bigl[ T^{*} \bigl( S^{1} \bigr)^{N} \Bigr] /{\CS_{N}}}
with the canonical symplectic structure
$$
\Omega_{I}^{red} = \sum_{i} \delta p_{i} \wedge \delta q_{i}
$$
The functions on $X_{I}$ invariant under the action
of $G$ descend down to $M_{I}$. Moreover the Poisson-commuting
functions descend to Poisson-commuting ones in the reduced
Poisson structure.
The Hamiltonian $H_{I}$ comes from the quadratic casimir
${\Tr} p^{2}$. The complete set of functionally independent
integrals is given by:
\eqn\inti{
H_{I}^{(k)} = {1\over{R^{k+1}}}{\Tr} p^{k+1}, \qquad k = 0, \dots, N-1}

\sssec{Construction \quad of \quad the \quad system \quad {\it II}}.
In the $II$ case one starts (again, essentially following \opr\kks )  with the symplectic
manifold $X_{II} = T^{*}\lieg \times {\IC}^{N}$ where $\lieg = Lie U(N)$
with the
canonical Liouville form
\eqn\cnfrmii{
\Omega_{II} = {\Tr}  \delta \bigl( P \wedge \delta Q  \bigr)
+ {1\over{2i}} \delta v^{+} \wedge \delta v}
The manifold $X_{II}$ is  acted on by $G$ (by conjugation
on the $T^{*}\lieg$ factor and in a standard way on $\IC^{N}$). The action
is Hamiltonian with the moment map:
\eqn\momii{
\mu_{II} = [P, Q] - v \otimes v^{+}}
One performs the reduction at the central level of the moment map, i.e.
takes the manifold $\mu_{II}^{-1} (- \nu \cdot {\Id} )$ and takes
its quotient by $G$ (as it is invariant). Explicitly, one solves the equation
\eqn\momiie{[P, Q] - v \otimes v^{+} = - \nu \Id}up to the $G$-action. The way to do it
is to fix a gauge
$$
Q = {\rm diag} (q_{1}, \ldots, q_{N})
$$
and then solve for $P$ and $v$. 
As a result one gets the reduced phase space:
\eqn\redii{
M_{I} = \Bigl[ T^{*} \IR^{N} \Bigr] /{\CS_{N}}}
The functions on $X_{II}$ invariant under the action
of $G$ descend down to $M_{II}$. Moreover the Poisson-commuting
functions descend to Poisson-commuting ones in the reduced
Poisson structure.
The Hamiltonian $H_{II}$ comes from the quadratic casimir
${\Tr} (P^{2} + \omega^{2}Q^{2}) $.
A convinient set of functionally independent
integrals is given by:
\eqn\intii{
H_{II}^{(k+1)} = {\Tr} (Z Z^{+})^{k+1}, \qquad k = 0, \dots, N-1}
with the matrix:
\eqn\hollax{
Z = P + i \omega Q \in \lieg_{\IC}}
The explicit form of $Z$, solving \momii\  is
$$
Z = {\rm diag}( p_{i} + i \omega q_{i} ) + \Vert  {{i{\nu}(1 - \delta_{ij})}\over{q_{i} - q_{j}}} \Vert
$$
In the paper \perlprep\ the matrix $Z$ was called $L^{+}$ and together with $L^{-} = Z^{\dagger}$
they formed a pair of "Lax-like" matrices. It was shown that $L^{\pm}$ evolve
according to the following generalization of Lax equations:
\eqn\gnlx{i {\dot L}^{\pm} = [ M, L^{\pm} ] \pm \omega L^{\pm}}
Clearly, this is just  the simple flow in the $Z, Z^{\dagger}$ space:
$$
Z(t) = e^{- i\omega t} Z(0) 
$$
projected onto the space of eigenvalues of $Q = {1\over{2i\omega}} (Z - Z^{\dagger})$.
See \perlbook\perelomov\ for the study of the relations between the trajectories
of the system $II$ and the $R \to \infty$ limit of the system $I$. 
\newsec{The Map $\pi$.}

\subsec{The Polar Decomposition}
The map $\pi$ is constructed first on the level of unreduced systems, and then 
by $G$- equivariance descends to the map of the reduced phase spaces. The map $\pi$ 
sends $v$ identically to $v$. Let us discuss its $Z \mapsto (p,g)$ part
(we also denote it by $\pi$).
The map $\pi$ sends $Z$ to its polar decomposition:
\eqn\mppi{\eqalign{& {\pi}(Z) = (p,g) \quad Z = {\omega}^{\half} p^{\half} g, \cr
& \quad g \in G, g^{\dagger}g = 1, p^{\dagger} = p, \quad p = {1\over{\omega}} ZZ^{\dagger}\cr}}
The image of ${\pi}$ is clearly the space $U$ of all pairs
$(p,g)$ where $p$ is a Hermitian matrix with non-negative eigenvalues, and
$g$ is a unitary matrix. It is a subset of $T^{*}G$, which we denote as
$T^{*}_{+}G$. Now it is a straightforward computation 
to check that 
\eqn\cnm{{\pi}^{*}\Omega_{II} = \Omega_{I}}
Clearly, $\pi$ commutes with the action of $G$:
\eqn\equ{{\pi}(UZ U^{-1}; U v) = U \cdot (p,g; v) \equiv ( U pU^{-1}, UgU^{-1}; U v) }
Finally, the Hamiltonians $H_{II}^{(k+1)} = {\Tr} (ZZ^{\dagger})^{k+1}$ map 
to $({\omega}R)^{k+1} H^{(k)}_{I} = {\Tr}({\omega}p)^{k+1}$.  So, in order to achieve
(a.) of  \mp\ one must choose
\eqn\rlt{{\omega} = {1\over{R}}}

Under the map ${\pi}$ the quadratic Hamiltonian $H_{I}$ of the Calogero  model is mapped to the linear 
(total momentum) Hamiltonian $H^{(0)}_{II} = \sum_{i} p_{i}$ of the Sutherland model. The quadratic
Hamiltonian of the Sutherland system is mapped to the quartic hamiltonian of
Calogero system.
\subsec{Generalizations}
One can obviously generalize the construction to cover the spin generalizations
of the Calogero \gibbons\ as well as of the Sutherland models (trigonometric case of \bbkt). In fact, the map
$\pi$ remains the same, the only change is that one starts with a bigger phase space:
$T^{*}\lieg \times S$, where the symplectic $G$-manifold $S$ produces
spin degrees of freedom (for example, one take as $S$ a general coadjoint
orbit of $G$, see \nhol\ for a discussion of a general spin system). The map $\pi$ is again an identity on $S$ and the same \mppi\ on 
$T^{*}\lieg$ mapping it to $T^{*}_{+}G$. 

\newsec{The Quantum Case}
The wave-functions of the Sutherland system can be described with the help
of a quantum version of the reduction we sketched in the previous section \poly\gncal\eq . 
In short, one fixes a representation $R$ of $G$. The simplest case is to take
$R = S^{N\nu}{\IC}^{N}$. Then one represents a wavefunction:
$$
\psi ( q_{1}, \ldots , q_{N}) 
$$
as a $R$-valued function $\Psi$ on $G$ with the following equivariance properties:
\eqn\eqiv{\Psi  (U g U^{-1}) = T_{R} (U) \Psi (g) }Using the known behavior of
$\Psi$ under the adjoint $G$ action one restricts
$\Psi$ onto the maximal torus $T$, where \eqiv\ implies that $\Psi$ takes values
in the $T$ - invariant subspace of $R$, which is one-dimensional for our choice
of $R$. Therefore one gets a number-valued function. In order to specify the wavefunction
uniquely (up to a phase factor) one uses all quantum Hamiltonians:
$\hat H^{(k)}_{I}$ and normalization conditions as well:
$$
\hat H^{(k)}_{I} \Psi_{\lambda}  = E^{(k)}_{I} ({\lambda}) \Psi_{\lambda}
$$ 
where $\lambda$ is a spectral parameter. 

Now the quantum version of a map $\pi$ looks as follows. Given a function 
$\Psi(g)$ obeying \eqiv\ one constructs a unique holomorphic 
function $\Psi (Z)$, which coincides with $\Psi (g)$ when $Z = g$. 
Then one passes to the real polarization, using the relations like:
$Z = {1\over{i}} ( {{\delta}\over{\delta Q}} - {\omega} Q)$ . 
One arrives at a function $\chi ( Q)$, which enjoys certain $G$-equivariance properties.
Therefore one can restrict $\chi$ to the diagonal matrices $Q = {\rm diag}(q_{1}, \ldots, q_{N})$
giving rise to the wavefunction of the Calogero system, which is the common 
eigenfunction of the family of Calogero quantum integrals  ${\hat H}_{II}^{(k)}$. The subtlelty in this transformation is that the quantum version of the operators:
$$
{\Tr} (ZZ^{\dagger})^{k+1}
$$ is defined up to the normal ordering ambiguitiy, which allows
one to shift ${\hat H}^{(k)}_{II}$ by an integral linear combination
of ${\hat H}^{(k^{\prime})}_{II}$ for $k^{\prime} < k$. In particular, the eigenvalues 
$E^{(1)}_{II} ({\lambda})$ may differ from $E^{(0)}_{I}({\lambda})$ by a $\lambda$
independent constant.

As an illustration of this, compare the formula for the spectrum of the Schr\"odinger operator
of the Calogero model (borrowed essentially from \c) and the trivial formula for the spectrum of
the total momentum operator of the Sutherland model. The formula of \c\ gives a spectrum of the operator
\eqn\clgr{{\hat H}_{0} = - {\half} \sum_{i=1}^{N} {{\p^{2}}\over{\p q_{i}^{2}}} + \sum_{i < j}
\Bigl( {{g}\over{q_{ij}^{2}}} + {{\omega_{0}^{2}}\over{4}} q_{ij}^{2} \Bigr)}
where $q_{ij} = q_{i} - q_{j}$ in the center-of-mass frame, i. e. on the set of functions,
annihilated by ${\hat P} = {1\over{i}} \sum_{i} {{\p}\over{\p q_{i}}}$. In order to map it to
the problem considered in this paper we use that 
\eqn\mdfcl{
{\hat H} = {\hat H}_{0} + N {\omega}^{2} {{q_{*}^{2}}\over{2}}, \quad q_{*} = {1\over{N}} \sum_{i=1}^{N} q_{i}
}
is the quantum counterpart of \hamii\ for 
$\omega = \omega_{0} \sqrt{N/2}$ (this model also has been considered  in \s). Notice, that the center-of-mass motion can be separated
(one can choose $q_{*}$ as one of coordinates, the rest of the coordinates being, say,
the Jacobi coordinates). We only need to know that  \c:
$$
- {\half} \sum_{i=1}^{N} {{\p^{2}}\over{\p q_{i}^{2}}} = -{1\over{2N}} {{\p^{2}}\over{\p q_{*}^{2}}} + \ldots
$$
to conclude that the spectrum of \mdfcl\ is the sum of the one of \clgr\ and $\omega ( {\half} + n)$, $n \in \IZ, n \geq 0$. Hence,
\eqn\spectrclgr{E^{Calogero}_{\vec n} = {\omega} \Bigl({{N}\over{2}} + {{N(N-1){\nu}}\over{2}} +\sum_{l=1}^{N} l n_{l}\Bigr), \quad n_{l} \geq 0} 
where $g = {\nu}({\nu}-1)$. This  spectrum can be represented as follows:
\eqn\egnvlscl{E^{Calogero}_{\vec n} = {\omega} \sum_{k=1}^{N} \bigl(  n_{k} + (k-1) {\nu} + {\half} \bigr), \quad 0 \leq n_{1} \leq \ldots \leq n_{N}} 
The integers $\vec n$ are  the spectral parameters $\lambda$ of the earlier discussion.

Now let us look at the Sutherland operator:
\eqn\sutherl{
{\hat H} = - {\half} \sum_{i=1}^{N} {{\p^{2}}\over{\p q_{i}^{2}}} + {{{\nu}({\nu}-1)}\over{4R^{2}}} \sum_{i > j} {1\over{{\rm sin}^{2}\bigl( {{q_{ij}}\over{2R}} \bigr)}}
 }
One can easily deduce from \s\ that the spectrum of this operator is given by:
\eqn\spectrsuth{E^{Sutherland}_{\vec n} = {1\over{2R^{2}}} \sum_{i} \bigl(n_{i} - {{\nu}\over{2}} (N+1- 2i) \bigr)^{2} \quad n_{1} \leq n_{2} \leq \ldots \leq n_{N} }
and the eigenvector with the eigenvalue $E_{\vec n}^{Sutherland}$ has the following form \s:
\eqn\egnfn{\Psi_{\vec n} ({\vec  q}) = \vert {\Delta} ({\vec q}) \vert^{\nu}\sum_{\vec n^{\prime}} c_{\vec n^{\prime}, \vec n} \Phi_{\vec n^{\prime}}}
where 
\eqn\orbsm{\Phi_{\vec n} = \sum_{\sigma \in {\CS}_{N}} e^{{{i}\over{R}} \sum_{k}  q_{k} n_{\sigma (k)}}, \quad {\Delta} ({\vec q}) = \prod_{i < j} {\rm sin}({{q_{ij}}\over{2R}}) }
The only important for us information about the coefficients $c_{\vec n^{\prime}, \vec n}$
is that for all $\vec n^{\prime}$ such that $c_{\vec n^{\prime}, \vec n} \neq 0$ the total momentum is conserved: $\sum_{k} n_{k}^{\prime}  = \sum_{k} n_{k}$. The total momentum eigenvalue is equal to:
\eqn\ttlmm{
P_{\vec n} = {1\over{R}} \sum_{k} \bigl ( (n_{k} -  {{\nu}\over{2}}(N-1)) + {\nu}(k-1) + {\half} \bigr) - {{N}\over{2R}} 
}
As a periodic function in $q_{k}$'s, $\Psi$ can be expanded in Fourier series.
The natural quantum analogue of the condition that the matrix $p \sim ZZ^{\dagger}$
has only positive eigenvalues is the condition that all Fourier modes of $\Psi$ are positive.
A little inspection of the structure of $\Delta ({\vec q})$ implies that all $n_{k}$
must obey the condition
$$
n_{k} \geq {{\nu}\over{2}}(N-1)
$$
Now it is easy to see that  up to the overall $\vec n$-independent shift \ttlmm\ and \egnvlscl\ actually do coincide, provided that 
$\omega R = 1$. Note in conclusion that in the quantum problem the coupling $g$ is written 
as $g = {\nu}({\nu}-1)$ while in the classical as ${\nu}^{2}$. As far as the relation between
the two systems is concerned the actual couplings $g$ are the same
$$
g_{I} = g_{II}
$$

\newsec{Discussion}

The phenomenon which we observed here is similar to a 
duality in integrable systems, recently discussed in \dual\. It is very interesting to see whether
this duality has anything to do with the relation between the $D$-brane matrix theories
and gauge theories on $D$-probes. Among more direct extensions of this work one may try to
find the analogue of the relation we pointed out for other root systems, as well as
for elliptic systems. For the latter purpose the existence of Hamiltonian reduction
\gnell\ may be of some use.

The Sutherland model was  called a system of Type II in Olshanetsky-Perelomov
classification of integrable pair potentials \op, while the Calogero model is of the Type V.
The systems I - III are  the degenerations of the elliptic system (type IV) with the potential $U(q)  = {\wp}(q)$. In that respect the system of the Type V was always a special case of such a  classification.
We have shown that it is actually a particular subsystem of the Type II model. 

Notice that one of the corollaries of our result is that the action-angle variables for the two systems coincide. Also, it implies that the system, dual to the Calogero model in the sense
of action-coordinate duality of \dual\ is nothing but the rational Ruijsenaars-Schneider model \rs\r.

\newsec{Acknowledgement}

I am grateful to Dima Ivanov for asking a question 
about the relation between the two models and to 
Sasha Gorsky and Volodya Rubtsov 
for the discussion which led to the results of the paper.

\listrefs
\bye